\documentclass[twocolumn,aps,floatfix,nofootinbib]{revtex4}
\usepackage{latexsym}
\usepackage{mathptm}
\usepackage{amsmath}
\usepackage{amssymb}
\usepackage{graphicx}
\def\beq{\begin{equation}}
\def\eeq{\end{equation}}
\def\beqn{\begin{eqnarray}}
\def\eeqn{\end{eqnarray}}
\newcommand{\f}{\begin{equation}}
\newcommand{\ff}{\end{equation}}

\begin{document}

\title{ The principle of relative locality  }
\author{Giovanni Amelino-Camelia$^a$, Laurent Freidel$^c$, Jerzy Kowalski-Glikman$^b$, Lee Smolin$^c$\thanks{lsmolin@perimeterinstitute.ca}
\\
$^a$Dipartimento di Fisica, Universit\`a ``La Sapienza" and Sez.~Roma1 INFN, P.le A. Moro 2, 00185 Roma, Italy\\
$^b$Institute for Theoretical Physics, University of Wroclaw,  Pl. Maxa Borna 9, 50-204 Wroclaw, Poland\\
$^c$Perimeter Institute for Theoretical Physics, \\ 31 Caroline Street North, Waterloo, Ontario N2J 2Y5, Canada
}

\date{\today}

\begin{abstract}
We propose a deepening of the relativity principle according to
which the  invariant arena for non-quantum physics is a phase space
rather than spacetime. Descriptions of particles propagating and
interacting in spacetimes are constructed by observers, but
different observers, separated from each other by translations,
construct different spacetime projections from the invariant phase
space.   Nonetheless, all observers agree that interactions are
local in the spacetime coordinates constructed by observers local to
them.

This framework, in which absolute locality is  replaced by relative
locality, results from deforming momentum space, just as the passage
from absolute to relative simultaneity results from deforming the
linear addition of velocities. Different aspects of momentum space
geometry, such as its curvature, torsion and non-metricity, are
reflected in different kinds of deformations of the energy-momentum
conservation laws.  These are in principle all measurable by
appropriate experiments.   We also discuss a natural set of physical
hypotheses which singles out the cases of momentum space with a
metric compatible connection and constant curvature.
\end{abstract}

\maketitle

%\tableofcontents

\section{Introduction}

How do we know we live in a spacetime? And if so how do we know we all share the same spacetime?
These are the fundamental questions we are investigating in this note.

We local observers do not
directly observe any events macroscopically displaced from our measuring
instruments.  As naive observers looking out at the world, and no
less as particle physicists and astronomers,
 we are basically ``calorimeters" with clocks.
 Our most fundamental measurements are
the energies and angles of the quanta we emit or absorb, and the
times of those events.  Judging by what we observe,  we live in
momentum space, not in spacetime.
%\footnote{Here and below by `momentum' we mean relativistic four-momentum.}

The idea that we live in a spacetime is constructed
by inference from our measurements of momenta and energy.  This was
vividly illustrated by Einstein's procedure to give spacetime
coordinates to distant events by exchanges of light signals \cite{AE-SR}.  When
we use Einstein's procedure we take into account the time it takes
the photons to travel back and forth but we throw away information about their
energy, resulting in a projection into spacetime. When we do this we
presume that the same spacetime is reconstructed by exchanges of
light signals of different frequencies.  We are also used to
assuming that different local observers, distant from each other,
reconstruct the same spacetime by measurements of photons they send
and receive.

But why should the information about the energy of the photons  we
use to probe spacetime be inessential?  Might that just be a low
energy approximation?  And why should we presume that we construct
the same spacetime from our observations as observers a cosmological
distance from us?

In this paper we show that absolute locality, which postulates that all observers live in the same space time,
is equivalent to the assumption that momentum space is a linear manifold.
This corresponds to an
 idealization in which we throw away
the  information about the energy of the quanta we use to probe
spacetime  and it can be transcended in a simple and powerful generalization of
special relativistic physics which is motivated by general
considerations of the unification of gravity with quantum physics.
In this work we link the notion of locality with assumptions made about the geometry of momentum space.
We propose a new framework in which we can relax in a controlled manner the concept of absolute locality
by linking this to a new understanding of the geometry of momentum space.
In this framework there is no notion of absolute locality, different observers see different spacetimes, and the spacetimes they observe
are energy and momentum dependent.
Locality, a coincidence of events,
becomes relative: coincidences of events are still objective for all
 local observers,
but they are not in general manifest in the
spacetime coordinates constructed by distant observers.

One way to motivate this new physical framework is by thinking about
the symmetry of the vacuum.   The most basic question that can be
asked of any physical system is what is the symmetry of the ground
state that governs its low lying excitations.   This is no less true
of spacetime itself, moreover in general relativity, and presumably
in any description of the quantum dynamics of spacetime, the
symmetry of the ground state is dynamically determined.  We also
expect that the classical spacetime geometry of general relativity
is a semiclassical approximation to a more fundamental quantum
geometry.  In this paper we show how simple physical assumptions
about the geometry of momentum space  may control the departure of
the spacetime description from the classical one.

We will first restrict attention to an approximation in  which
$\hbar$ and $G_{Newton}$ both may be neglected while their ratio
\f
\sqrt{\frac{\hbar}{G_{Newton}}} = m_p
\label{regime}
\ff
is held fixed\footnote{We
work in units in which $c=1$.}.   In this approximation quantum and
gravitational effects may both be neglected, but there may be new
phenomena on scales of momentum or energy given by $m_p$.  At the
same time, because $l_p = \sqrt{\hbar G_{Newton}} \rightarrow 0$ we
expect no features of quantum spacetime geometry to be relevant.

Since our approximation gives us an energy scale, but not a length
scale,  we will begin by presuming that momentum space is more
fundamental than spacetime.  This is in accord with the operational
point of view we mentioned in the opening paragraph.  So we begin in
momentum space by asking how it may be deformed in a way
that is measured by a scale $m_p$.  Once that is established we will
derive the properties of spacetime from dynamics formulated in
momentum space.  For convenience we work in first in the limit just
described, after which we will briefly turn on $\hbar$.

By following this logic below, we will find that physics may be
governed by a novel principle, which we call the  {\it Principle of
Relative Locality.} This states that,\newline

{\it Physics takes place in phase space and  there is no invariant
global projection that gives a description of processes in
spacetime. From their measurements local observers can construct
descriptions of particles moving and interacting in a spacetime, but
different observers construct different spacetimes, which are
observer-dependent slices of phase space.}\newline

%We postpone to the closing remarks of this paper a more careful
% discussion of the relationship between the relativistic framework we here advocate  and previous ones. see earlier draft

In the next section we introduce an operational approach to the
geometry of momentum space, which we build on in section III to give a
dynamics of particles on a curved momentum space.  We see how a
modified version of spacetime geometry is emergent from the dynamics
which is formulated on momentum space.  In these sections we
consider a general momentum space geometry, which illuminates a
variety of new phenomena that might be experimentally probed
corresponding to the curvature, torsion and non-metricity of
momentum space. However, one advantage of this approach is that with a
few reasonable physical principles the geometry of momentum space
can be reduced to three choices, depending on the sign of a parameter.
As we show in section IV, this
gives this framework both great elegance and experimental
specificity. In section V we make some preliminary observations as
to how the geometry of momentum space may be probed experimentally,
after which we conclude.

\section{An operational approach to the geometry of momentum space}

We take an operational point of view in which we describe physics
from the point of view of a local observer who is equipped with
devices to measure the energy and momenta of elementary particles in
her vicinity.  The observer also has a clock that measures local
proper time.  We construct the geometry of  momentum
space\footnote{By which we mean the space of relativistic
four-momenta denoted $p_{a}$ with $a=0,1,2,3$.}  $\cal P $ from measurements made of the dynamics of
interacting particles.   We assume that to each choice of
calorimeter and other instruments carried by our observer there is a preferred
coordinate on momentum space, $k_a$.
But we also assume that the dynamics can be expressed covariantly in
terms of geometry of $\cal P$ and do not depend on the choice of calorimeter's coordinates.
  We note that the $k_a$ measure the
energy and momenta of excitations above the ground state, hence the
origin of momentum space, $k_a=0$,  is physically well defined.

Our local observer can make two kinds of measurements.
One type of measurement can be done only with a single particle and it defines, as we will see, {\it a metric} on momentum space.
The other type of measurement involve multi particles and defines {\it a connection}. A key mathematical idea underlying our construction is that a connection on a manifold can be determined by an algebra \cite{Laurent-math}, in the present case this will be an algebra that determines how momenta combine when particles interact.

\subsection{The metric geometry of momentum space}

  First we describe the metric geometry. Our local observer can measure the rest energy or relativistic mass of a particle which is a function  of the four momenta.
    She can also measure the kinetic energy $K$ of a particle of mass $m$ moving with respect to her, but local to her.
  We postulate that these measurements determine the metric geometry of momentum space.
   We interpret the mass as the geodesic distance from the origin, this gives the dispersion relation
   \f
D^{2}(p)\equiv D^{2}(p,0)=m^{2}.
  \ff
  The measurement of kinetic energy defines the geodesic distance between a particle $p$ at rest and a particle $p'$ of identical mass and kinetic energy $K$, that is $D(p)=D(p')=m$ and
  \f
  D^{2}(p,p') = -2 m K.
  \ff
The minus sign express the fact that the geometry of momentum space is Lorentzian.
From these measurements one can reconstruct a metric on $\cal P$\footnote{In the standard case of physics in Minkowski
spacetime, $h^{ab}$ is the dual Minkowski metric and ${\cal K}_a (k)
=\sum_I k^I_a$.  A scale $m_p$ may be introduced by deforming the
geometry of $\cal P$ so that it is curved. The correspondence
principle (to be introduced below) assures that we recover the standard flat geometry of
$\cal P$ in the limit $m_p\rightarrow\infty$.}
 \f
 dk^2 =  h^{ab} (k) dk_a d k_b\label{metricP}.
\ff

\subsection{The algebra of interactions}

Now we describe the construction of the connection on momentum space.
This is determined by processes in which $n$ particles
interact, $n_{in}$ incoming and $n_{out}$ outgoing, with
$n=n_{in}+n_{out}$.  This
proceeds by the construction of an algebra, which then determines the connection.

Associated to each interaction there  must be a
combination rule for momentum, which will in general be
non-linear.   We denote this rule for two particles by
\f
(p, q)
\rightarrow p^\prime_a = ( p \oplus q )_a
\ff
Hence the momentum
space $\cal P$ has the structure of an algebra defined by the
product rule $\oplus$.  We assume that more complicated processes
are built up by iterations of this product -- but to begin
with we assume neither linearity, nor commutativity nor associativity.

We will also need an operation that turns
outgoing momenta into incoming momenta.  This is denoted, $p
\rightarrow \ominus p$ and it satisfies\footnote{And more generally
$(\ominus p )\oplus (p\oplus k)= k$, where  $\ominus$ is a left inverse. }
\f
 ( \ominus p )\oplus p =0
\ff

Then we have the conservation law of energy  and momentum for any
process, giving, for each type of interaction, four functions on
${\cal P}^n$, depending on momenta of interacting particles, which
vanish \f {\cal K}_a (k^I) =0 \label{conserve1} \ff For example, for
a process with three incoming momenta $p_a$, $q_a$, and $k_a$ one
might make the choice
\begin{equation}\label{8}
    {\cal K}_a (p,q,k) =p_a\oplus(q_a\oplus k_a)=0
\end{equation}

\subsection{From the algebra of interactions to the connection on momentum space}

Corresponding to the algebra of combinations of momenta there is  a
connection on $\cal P$.  The geometry of momentum space is studied
in detail in \cite{Laurent-math}, but the basics are as
follows. The algebra of the
combination rule determines a connection on $\cal P$ by \f
\frac{\partial}{\partial p_a}\frac{\partial}{\partial q_b} (p \oplus
q )_c |_{q,p=o} = -\Gamma_c^{ab} (0) \label{connection1} \ff The
torsion of $\Gamma_a^{bc}$ is a measure of the asymmetric part of
the combination rule \f -\frac{\partial}{\partial
p_a}\frac{\partial}{\partial q_b} \left ( (p \oplus q )_c  -  (p
\oplus q )_c \right )_{q,p=o} = T_c^{ab} (0) \label{torsion1} \ff
Similarly the curvature of $\cal P$ is a measure of the lack of
associativity of the combination rule

 \f
 2\frac{\partial}{\partial
p_{[a}} \frac{\partial}{\partial q_{b]}} \frac{\partial}{\partial k_c}
\left (  (p \oplus q ) \oplus k   -  p \oplus (q \oplus k )   \right
)_d |_{q,p,k=o} = R_{\ \ \  d}^{abc} (0) \label{curvature1}
\ff
where the bracket denote the anti-symmetrisation.
%Notice that connection, torsion, and curvature vanish in the limit $m_p\rightarrow\infty$.

We note that there is no physical reason to expect a combination
rule  for momentum to be associative, once it is non-linear.
Indeed, the lack of associativity means there is a physical
distinction between the two processes of Figure 1, which is
equivalent to saying there is a definite microscopic causal
structure. That is, {\it  causal structure of the physics maps to
non-associativity of the combination rule for momentum which in turn
maps to curvature of momentum space.}  The curvature of momentum
space makes microscopic causal orders distinguishable, and hence
meaningful.  This gives rise to proposals to measure the curvature
of momentum space which we will discuss below.

\begin{figure}[hbt]
\begin{center}
\includegraphics[width=0.5 \textwidth]{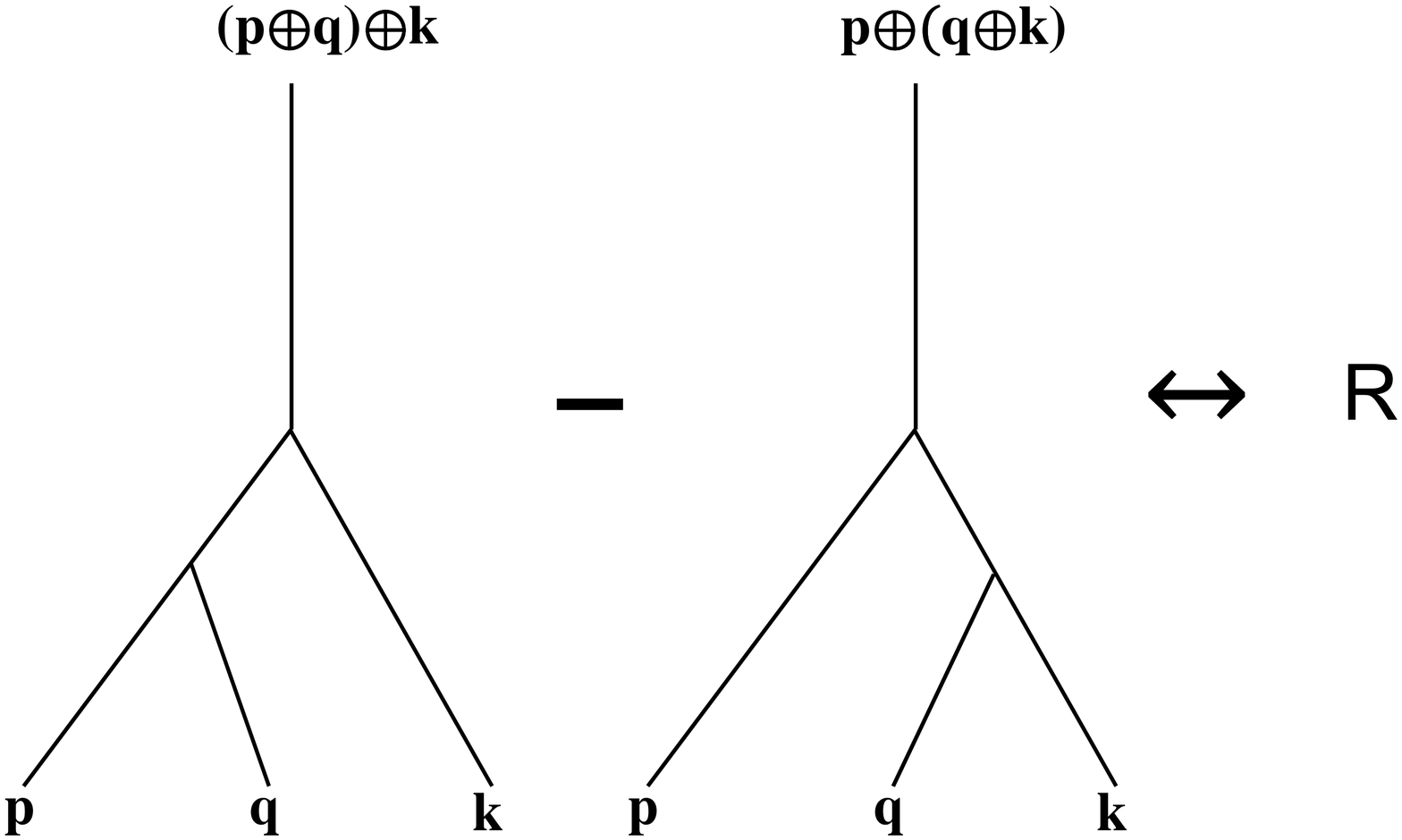}
\end{center}
\caption{Curvature of the connection on momentum space produces non-associativity of composition rule.}
\end{figure}

To determine the connection, torsion and curvature away from the
origin of momentum space we have to consider translating in momentum
space, ie we can denote
\f \label{transprod}
p \oplus_k q = k \oplus ((\ominus k \oplus p) \oplus (\ominus k \oplus q)) \ff
the identity for this product is  at $0_{k}=k$.
Then \f \frac{\partial}{\partial p_a}\frac{\partial}{\partial q_b}
(p \oplus_k q )_c |_{q,p=k} = -\Gamma_c^{ab} (k) \label{connection2}
\ff
 Thus, the  action of adding an infinitesimal momentum $dq_a$
from particle $J$ to a finite momentum $p_a$ of particle $I$ defines
a parallel transport on $\cal P$.
\f p_a \oplus dq_a = p_a + dq_b
\tau_{a}^{b}(p)\label{conserve22}
\ff
where $\tau(p)$ is the parallel transport operation from the identity to $p$.
It can be expanded around $p=0$
\f
\tau_{a}^{b}(p) = \delta_{a}^{b} -\Gamma_{a}^{bc}p_{c} - \Gamma_{a}^{bcd}p_{c}p_{d} +
\cdots
\label{conserve2}
\ff
with
\f
\Gamma_{a}^{bcd} = \partial^{d}_{p} \Gamma_{a}^{bc} - \Gamma_{i}^{db}\Gamma_{a}^{ic}- \Gamma_{i}^{dc}\Gamma_{a}^{bi}.
\ff
 The corresponding conservation law thus has the form to second order
\f {\cal K}_a (k) =\sum_I  k^I_a - \sum_{  J \in {\cal J}(I)}
C_{I,J} \Gamma_a^{bc} k_b^J k_c^I +... \label{conserve3} \ff where $
{\cal J}(I)$ is the set of particles that interact with the I's one
and $C_{I,J}$ are coefficients that depend on the form of the
conservation law.

We will shortly study the consequences of curvature and torsion on
momentum space for the dynamics of particles. We will see that the
meaning of the curvature of momentum space is that  it implies a
limitation of the usefulness of the notion that processes happen in
an invariant spacetime, rather than in phase space.  The hypothesis
of a shared, invariant spacetime, in which all observers agree on
the locality of distant interactions, turns out to be a direct
consequence of the linearity of the usual conservation laws of
energy and momentum.   When we deform the conservation laws by
making them non-linear, this gives rise to relative locality.

\section{The emergence of spacetime from the dynamics of particles}

We take the point of view that spacetime is an auxiliary concept which emerges when we seek to define dynamics in momentum space.
 If we take the momenta of elementary particles to be primary, they themselves need momenta, so that a canonical dynamics can be formulated. The momenta of the momenta are quantities
$x^a_I$ that live in the cotangent space of ${\cal P}^n$ at a point $k_a^I$.

\subsection{Variational principle}

Given these we can define the free particle dynamics by
%\footnote{One can consider more general actions for free particle, see eg \cite{Arzano:2010kz} and references therein. Here we will consider only the simplest option.}
\f
S^I_{free}= \int ds \left ( x^a_J \dot{k}_a^J + { \cal N}_J {\cal C}^J (k)
\right )
\label{free}
\ff
where $s$ is an arbitrary time parameter and ${\cal N}_{J}$ is the Lagrange multiplier imposing the mass shell condition
\f\label{ep1}
{\cal C}^{J}(k)\equiv D^{2}(k) -m^{2}_{J}.
\ff
 We emphasize that the contraction $x^a_J \dot{k}_a^J$ does not involve a metric, and the dynamics is otherwise given by constraints which are functions only of coordinates on $\cal P$ and depend only the geometry of $\cal P$.    This leads to the Poisson brackets,
\f
\{ x^a_I , k_b^J \} = \delta^a_b \delta_I^J
\ff

We then have a phase space, $\Gamma$ of a single particle, which is the cotangent bundle of $\cal P$.  We note that there is neither an invariant projection from $\Gamma$ to a spacetime, $\cal M$ nor is there defined any invariant spacetime metric.  Yet this structure is sufficient to describe the dynamics of free particles.  The fact that there is no invariant projection to a spacetime is related to the non linearity of momentum space.    Indeed under a non linear
redefinition $ p \to F(p)$ the conjugated coordinates is given by $x \to (\partial p /\partial F ) x$, so the new canonical coordinate appears to be momentum space dependent. This is this mixing between ``spacetime'' and momentum space that is the basis of the relative locality.
We can call the $x^a_J$ {\it Hamiltonian spacetime coordinates} because they are defined as being canonically conjugate to coordinates on momentum space.

Note that we  do not need a spacetime or spacetime metric to describe how these particles interact.  If we consider the process with $n$ interacting particles we want to impose conservation of the non-linear quantities, ${\cal K}_a$.  We do this by introducing a lagrange multiplier to guarantee conservation (\ref{conserve1}).  The action is
\f
S^{total}= \sum_J S^J_{free} + S^{int}\label{actiontotal}
\ff
where for the incoming particles
\f
S^J_{free}= \int_{- \infty}^0  ds \left( x^a_J \dot{k}_a^J +  {\cal N}_J {\cal C}^J (k)
\right)
\label{free-in}
\ff
while for the outgoing particles
\f
S^J_{free}= \int_{0}^\infty  ds \left ( x^a_J \dot{k}_a^J + {\cal N}_J {\cal C}^J (k)
\right )
\label{free-out}
\ff
The interaction contribution to the action is simply a lagrange multiplier times the conservation law  (\ref{conserve1}).
\f
S^{int}= {\cal K}(k(o))_a z^a \label{interactionS}
\ff

We have set the interaction to take place at affine parameter $s=0$ for each of the particles.  At this point
$z^a$ can be just considered to be a lagrange multiplier to enforce the conservation of momentum (\ref{conserve1}) at the interaction where for each particle
$s=0$.

We vary the total action.  After an integration by parts in each of the free actions we have
$$
\delta S^{total} =$$\f \sum_J \int_{s_1}^{s_2} \left (  \delta x_J^a \dot{k}_a^J - \delta k_a^J \left[\dot{x}^a_J - {\cal N}_J \frac{\delta {\cal C}^J}{\delta k_a^J} \right]
+\delta {\cal N}_J {\cal C}^J (k)
\right )   + {\cal R}
\ff
Here $\cal R$ contains both the result of varying $S^{int}$ and the boundary terms from the integration by parts.  $s_{1,2}$ are $0,\infty, -\infty$ depending on whether the term is incoming or outgoing.  Before examining the boundary terms we confirm we have the desired free parts of the equations of motion
\begin{eqnarray}
\dot{k}_a^J &=& 0
\nonumber \\
\dot{x}^a_J &=& {\cal N}_J \frac{\delta {\cal C}^J}{\delta k_a^J}
\nonumber \\
{\cal C}^J (k) &=&0
\label{EOM1}
\end{eqnarray}
We fix $\delta k_a^J =0$ at $s=\pm \infty$ and examine the remainder of the variation
\f
{\cal R}= {\cal K}(k)_a \delta z^a - \left (  x^a_J (0) - z^b  \frac{\delta {\cal K}_b}{\delta k_a^J }  \right )  \delta k_a^J
\ff
Here $x^a_J$ and $k_a^J$ are taken for each particle at the parameter time $s=0$.
This has to vanish if the variational principle is to have solutions.  From the vanishing of the coefficient of $\delta z^a$ we get the four conservation laws
of the interaction (\ref{conserve1}).  From the vanishing of the coefficient of $\delta k_a^J$ we find $4n$ conditions which hold at the interaction
\f
 x^a_J (0) =  z^b  \frac{\delta {\cal K}_b}{\delta k_a^J }
\label{interact} \ff Using (\ref{conserve3}) this gives conditions
\f x^a_J (0) = z^a  - z^b \sum_{  L \in {\cal J}(J)} C_{J,L}
\Gamma_b^{ac }   k_c^L+... \label{interact1} \ff This tells us that
to leading order, in which we ignore the curvature of momentum
space, all of the worldlines involved in the interaction meet at a
single spacetime event, $z^a$.  The choice of $z^a$  is not
constrained and cannot be, for its variation gives the conservation
laws (\ref{conserve1}). Thus, we have recovered the usual notion
that interactions of particles take place at single events in
spacetime from the conservation of energy and momentum. This is good
because in quantum field theory conservation implies locality, and
it is good to have a formulation of classical interactions where
this is also the case.

However when we include terms proportional to $z^a$, which is to say
when the observer is not at the interaction event, we see that the
relationship between conservation of energy and momentum and
locality of interactions is realized a bit more subtly. The
interaction takes place when the condition (\ref{interact1}) is
satisfied, that is at $n$ separate events, separated from $z^a$ by
intervals \f \Delta x^a_J (0) = - z^b \sum_{  L \in {\cal J}(J)}
C_{J,L}\Gamma_b^{ac }   k_c^L +  ... \label{interact2} \ff

These relations (\ref{interact1}), (\ref{interact2}) illustrate concisely the relativity of locality.  For some fortunate observers
the interaction takes place at the origin of their coordinates, so that $z^a= x^a_J(0)=0$ in which case the interaction is observed to be local.
Any other observer, translated with respect to these, has a non-vanishing $z^a$ and hence sees the interaction to take place at a distant set of
events.  These are centered around $z^a$ but are not precisely at the same values of the coordinates.  That is the coordinates of particles involved
in an interaction removed  from the origin of the observer by a vector $z^a$ are spread over a region of order
\f
\Delta x \approx |z| |\Gamma | k
\label{uncertain1}
\ff
The relationship (\ref{interact}) possess a very nice mathematical meaning too.
Since the momentum space is in general curved the proper way to define the conjugate coordinates is as elements of the cotangent bundle of $\cal P$.
The cotangent space based at $p^{I}$ and the cotangent space based at $0$ are different spaces in the general curved case.
This expresses mathematically the relativity of locality.
The hamiltonian particle coordinate $x_{I}$ represent an element of $T^{*}_{p^{I}}{\cal P}$ while the interaction coordinate being dual to
the conservation law represent an element of $T^{*}_{0}{\cal P}$.  (\ref{interact}) represent a relation between these two spaces and remarkably it can be shown
\cite{Laurent-math} that indeed the coefficient $\partial{\cal K}/\partial k$ evaluated when ${\cal K}=0$ is the parallel transport operator, more precisely
\f
\partial_{p}^{b}(k\oplus p)_{a}|_{p=\ominus k} = (\tau(k)^{-1})_{a}^{b}
\ff
where $\tau(k)$ is the parallel transport operator of vectors from $0$ to $k$ introduced earlier
therefore $\tau(k)^{-1}$ is the parallel transport operator of covectors from $0$ to $k$.

\subsection{The physical meaning of relative locality}

Is this a real, physical non-locality or a new kind of coordinate artifact?   It is straightforward to see that it is the latter, because the $\Delta x^a_J (0) $ can be made to vanish by making a translation to the coordinates of another observer.  In a canonical formulation, translations are generated by the laws of conservation of energy and momentum, (\ref{conserve1}).  Given any local observable in phase space ${\cal O}$ observed by a local observer, Alice, we can construct the observable as seen in coordinates constructed by another observer, Bob, distant
from Alice, by a translation labeled by parameters $b^a$.
\f
\delta_b {\cal O} = b^b \{ {\cal K}_b ,  {\cal O} \}
\label{translate1}
\ff
Since momentum space is curved, and $ {\cal K}_b $ is non-linear, it follows that the ``spacetime coordinates" $x^a_J$ of a particle translate
{%\underline
{in a way that is dependent
on the energy}}
{%\underline
{and momenta of the particles it interacts with}}, $x^a_J
\rightarrow x^{\prime a}_J(0)= x^a_J (0) +  \delta_b x^a_J (0)$
where \f
 \delta_b x^a_J (0) = b^b \{ {\cal K}_b , x^a_J \}  =-  b^a +    b^b
 \sum_{  L \in {\cal J}(J)} C_{J,L}\Gamma_b^{ac }   k_c^L+  ...
\label{translate2} \ff This is a manifestation of the relativity of
locality, ie local spacetime coordinates for one observer mix up
with energy and momenta on translation to the coordinates of a
distant observer.

This mixing under translations effect also entirely accounts for the separation of an interaction into apparently distinct events, because
with $b^b=-z^b$, we see that $\Delta x^a_J$ of (\ref{interact}) is equal to $ \delta_b x^a_J (0) $ of (\ref{translate2}).  Thus, the observer whose new coordinates we have translated to observes a single interation taking place at $x^a_J \rightarrow x^{\prime a}_J(0)=0$.

Thus, if I am a local observer and see an interaction to take place via a collision at my origin of coordinates, a distant observer will generally see it in their coordinates as spread out in space-time by (\ref{interact}).  And vice versa.  There is not a physical non-locality, as all momentum conserving interactions are seen as happening at a single spacetime event by some family of observers, who are local to the interaction.
But it becomes impossible to localize  distant interactions in an absolute manner.
%What is invariant is the description in phase space, what is relative is how spacetime coordinates are projected out of the invariant phase space.
Furthermore, all observers related by a translation agree about the momenta of the particles in the interaction, because under translations (\ref{translate1})
$\delta_b k_a^I =0$.

Note that if the curvature and torsion vanish there is no mixing of spacetime coordinates with momenta under translations, so there is an invariant definition of spacetime. Thus, the flatness of momentum space is responsible for the notion of an absolute spacetime, just as the additivity of velocity allows Newtonian physics to have an absolute time.

Note also that the translations of spacetime coordinates define
sections $\cal S$ on $\Gamma$, which extend from the origin of local
coordinates. These tell us how to translate events at the origin of
the coordinates of an observer to coordinates measured by a distant
observer. These sections provide local and energy dependent
definitions of space-time, relative to observers and energy scales.
These sections are defined by Hamiltonian vector fields on $\Gamma$
which are defined acting on functions $f$ on $\Gamma$ by \f v_b f =
\{ b^a {\cal K}_a , f \} = b^a \sum_J  \left (1  - \sum_{ K \in
{\cal J}(J)}C_{J,K} \Gamma_a^{bc} k_c^K  +...  \right )
\frac{\partial f}{\partial x^a_J} \ff We can check that these
commute and hence define submanifolds of $\Gamma$.   We can define
an inverse metric on the sections $\cal S$ defined by \f g^{ab} (x,
k) = g ( dx^a , dx^b ) = h^{cd}(k) \{ {\cal K}_c , x^a \} \{{\cal
K}_d, x^b \} \label{gdef1} \ff

We note that this metric on the sections $\cal S$ is momentum dependent.  Thus, we arrive at a description of the geometry of spacetime which is energy dependent.
This metric is in fact just the fiber metric at the point $k$ where the fiber is the cotangent plane at this point\footnote{ We are usually familiar with such a phenomenon in the dual picture where gravity is turned on and  spacetime is curved, in which case momentum space is represented by covector fields and the metric induced on each fiber is dependent on the spacetime point. It can be said that relative locality is a dual gravity. }.

But note that we can take all the $k_a^I=0$ in which case $ g^{ab} (x, k=0)  = \eta^{cd}$ is the Minkowski metric. So observers who probe spacetime with zero momentum probes will see Minkowski spacetime.  However, the coordinates of this invariant zero momentum section are non-commutative.
\f
\{ z^a , z^b \} \neq 0
\ff
Indeed, if one wants to describe the spacetime as probed by the zero momentum probes this means that
we desire to model the spacetime as  the cotangent space of the origin. In order to achieve this we need to parallel transport  the
event at $p$ back to events at $0$.
This means that we interpret the coordinates $z$ as being covariant covector fields defined by
\f
z^{a} = \tau(p)_{b}^{a} x^{b}_{p}
\ff
where $x_{p}$ is the coordinate dual to $p$ with respect to the poisson bracket and
living in the cotangent space at $p$.
The Poisson commutator can now be evaluated, it is related to the Lie bracket of the covariantly constant vector field on $\cal{P}$ and its expansion is given by
\begin{eqnarray}
\{ z^a , z^b \} &=&( \tau^{a}_{\bar{a}} \partial^{\bar{a}} \tau^{b}_{\bar{b}} - \tau^{b}_{\bar{a}} \partial^{\bar{a}} \tau^{a}_{\bar{b}}) x^{\bar{b}}\nonumber \\
&=& T^{ab}_{d} z^{d} + R^{abc}{}_{d}p_{c}z^{d} +\cdots
\end{eqnarray}
where we have expanded around $p=0$ in the second equality.

However, when we go to zero momentum we can no longer neglect the limitations on local measurements coming from the uncertainty principle of quantum mechanics.
Quantum mechanically a particle of energy $p_0$ can only be localized with accuracy
 not greater than $\hbar /p_0$, and this combines with
the relativity of locality expressed by (\ref{uncertain1}) to give uncertainty relation of the form
\f
\Delta x \geq \frac{\hbar}{ p_0} + |x| |\Gamma | p_0 ~,
\ff
 characterizing the limitation on the sharpness of coordinates of particles with
energy $p_0$. Taking small $p_0$ helps reduce the relative-locality features
but increases the quantum mechanical uncertainty.

To understand the implications of this in more detail we will next specialize from the general case by imposing physical conditions which restrict the geometry of momentum space.

\section{Specializing the geometry}

As we have seen, the geometry of momentum space can code several kinds of deformations of the energy-momentum conservation
laws, which take advantage of the flexibility to choose the metric, torsion, curvature and non-metricity of the connection.
This gives us an arena within which we can formulate and test
new physical principles.  These impose constraints on the choice of the geometry of $\cal P$. To illustrate
this we next turn from the general case to show how a set of simple
principles restricts us to a one parameter set
of momentum space geometries, and consequently, an almost unique set of experimental predictions.

Consider the following four increasingly strong physical principles:

\begin{enumerate}

\item{\it The correspondence principle}:   Special relativity describes accurately
all processes involving momenta small compared to some mass scale $m_c$. While it is natural to presume that $m_c \approx m_p$, the scale
$m_c$ should be determined experimentally.

\item{}{\it The weak  dual equivalence principle: }   The algebra of combination of momenta,
and hence the geometry of $\cal P$ are universal; they do not depend on which kinds of particles are involved in interactions.

\item{\it The strong dual equivalence principle or $E=mc^{2}$:} There is equivalence between the rest mass energy defined by the metric and the inertial mass
which involves the connection.

\item{}{\it Maximal symmetry.}
The geometry of momentum space is isotropic and homogeneous: invariant under Lorentz transformations and invariant under ``translations''.

\end{enumerate}

The geometry of momentum space is discussed in more detail in
\cite{Laurent-math} where it is precisely shown  how these principles
lead to a unique geometry. This unique geometry is characterized by a constant: the dual cosmological constant which as the
dimension of an inverse mass square. When this dual cosmological constant is zero we recover usual special relativity, when it is non zero,
 momentum space is genuinely curved.

 The first principle
implies first that the metric of momentum space is a Lorentzian metric ( which we already have implicitly assumed).
It also implies that the torsion and non-metricity of $\Gamma_a^{bc}$ must
be at least of order ${1}/{m_p}$, while curvature
%\footnote{Notice that we cannot impose
%restrictions on the symmetric part of $\Gamma_a^{bc}$. The reason is
%that even in special relativity with flat momentum space we can
%introduce non-standard momentum variables, being nonlinear functions
%of the usual ones (for more details see \cite{MicheleJurek}). In
%terms of these new variables both the on shell condition and the
%momentum composition rule would change, and the metric part of
%$\Gamma_a^{bc}$ will be nontrivial. However, such a change of
%variables clearly does not change physics, since it corresponds to
%regauging of the calorimeter.}
 must be of
order of ${1}/{m_p^2}$.

The weak equivalence principle implies that the combination of momenta do not depend on the colors or charges of particles and is the same
as the composition for identical particles. For identical particles there is no operational way to give an order of the combination rule
if we have Bose statistics, therefore  taken strongly, that is if we do not allow for any modification of the statistics of identical particles, this principle also implies that the product is symmetric and hence the connection is torsionless.

The strong equivalence principle relates the metric and the
connection of $\cal P$ by imposing that the connection is metric
compatible. The metric determines distance between two points in
momentum space, and hence governs the mass shell relations of single
particles, while the?connection determine what is the straightest
path between two points, and?hence is determined by interactions
which combine momenta. Since they are given by different physics,
they are in principle independent. However, there are indications,
to be discussed in \cite{E=MC2} that, at least in some cases, the
 non-metricity of the connection is related to violations of the  equivalence between relativistic energy and mass.  It is intriguing that Einstein's
 observation that in a relativistic theory $E=mc^2$ appears  to relate the metric and the connection of momentum space.

The first three principles impose therefore that the geometry of
momentum space  is entirely fixed by a Lorentzian metric. The
connection is then the unique connection which is torsionless and
compatible with the metric.

The fourth principle of maximal symmetry is the most restrictive.
 This could be called the principle of ''special relative locality'' since it essentially implies a unique fixed dual geometry on momentum space.
What it means in a nutshell is that the space of Killing vector
fields of the metric form a ten dimensional Lie algebra. This
symmetry algebra also preserve the connection. This implies that
there exists Lorentz transformations   $\Lambda$ acting on $\cal P$
fixing the identity $0$ such that \f \Lambda(p\oplus q)
=\Lambda(p)\oplus \Lambda(q). \ff This implies that the conservation
law transform covariantly under Lorentz transformations \f {\cal
K}(\Lambda k) =\Lambda {\cal K}(k). \ff

Hence,  if we impose that the geometry of $\cal P$ is invariant
under Lorentz  transformations then we gain an action of the Lorentz
group on the phase space $\Gamma$.   From this we can conclude that
there are for each interaction event, a family of local observers
which see the interaction to take place at $z^a=0$ and hence be
local.

There also exist a notion of translations\footnote{The algebra of
translations  on momentum space does not have to be commutative. It
can be  defined to be the left translation $T_{r}(p)\equiv r \oplus
p$. } $T_{r}$ such that $T_{r}(0)=r$ and \f T_{r}(p\oplus_{k} q) =
T_{r}(p)\oplus_{T_{r}k} T_{r}(q) \ff where $\oplus_{k}$ is the
translated combination rule (\ref{transprod}).

Here we expand more on how Lorentz invariance can be made compatible
with curved momentum space. First it is well known that around the
identity $0$ we can set up a special set of coordinates:  The
Riemannian normal coordinates. In this coordinates the distance from
$0$ is given by the usual flat space formula hence the mass shell
condition in this coordinates is simply\footnote{The distance
$D^{2}(k,k')$ of two points away from the origin do not assume this
simple form (\ref{simpleform}), which applies only to measurements
of distances from the origin.  Consequently, the action of lorentz
boosts on the geometry can be non-trivial even if the action on the
coordinates $k_a$ in which  (\ref{simpleform}) holds is linear.} \f
{\cal C}(k)= k_{0}^{2} - k_{i}^{2} - m^{2} =0. \label{simpleform}
\ff The Lorentz transformations that preserves the zero momenta and
the metric therefore acts in the usual manner in this coordinate
systems.

Moreover under the hypothesis of homogeneity these coordinates can
be extended to cover  almost all the manifold $\cal P$. The Lorentz
generators therefore satisfy the usual algebra. If we assume in turn
that the Lorentz transformations are canonical transformations
preserving the Poisson bracket, they also satisfy the usual Poisson
algebra.
 That is given the boost and rotation generators $N_i$ and $M_j$, $i,j,k =
1,2,3$,  we have
\begin{equation}\label{j1}
    \{M_i, M_j\} = \epsilon_{ijk}\, M_k, \quad \{M_i, N_j\} = \epsilon_{ijk}\, N_k, \quad \{N_i, N_j\} = -\epsilon_{ijk}\, N_k
\end{equation}
and the generators act on the momentum space $\cal P$ through these
brackets.  Moreover, as in special relativity, we assume that  the
momentum space $\cal P$ splits into collections of orbits of the
Lorentz group: the zero momentum point, which is left invariant by
Lorentz transformations; the positive and negative energy mass
shells of massive particles; the light cone corresponding to
massless particles; and mass shells of tachyons of imaginary mass.
It is a direct consequence of this assumption that the function
${\cal C}(k)$ in the mass shell condition (\ref{ep1}) must be a
Lorentz scalar, so that all Lorentz observers agree what the value
of the invariant mass $m^2$ is. It follows then that vectors
corresponding to the infinitesimal Lorentz transformations are
Killing vectors of the metric (\ref{metricP}), so that the surfaces
of constant distance from the origin (the point in $\cal P$
corresponding to zero momentum) are orbits of action of Lorentz
group.

Combining the assumptions of Lorentz symmetry with translation
invariance or, equivalently, homogeneity then completely determines
the geometry of $\cal P$, up to an overall scale. Indeed, according
to the latter the geometry of a fixed mass orbit $|k|^2+m^2=0$ is
the same of all the masses. Thus the geometry of $\cal P$ is not
only invariant under Lorentz transformations acting along the
orbits, but also under translations, mapping one orbit to another.
All together we have therefore the 10-parameter (in 4 dimensions)
group of transformations that leave the geometry of the momentum
space invariant, and therefore this space is a maximally symmetric
manifold. It is well known that there are only three such manifolds:
the flat space and the (anti-) de Sitter spaces. On all three of
these spaces the Lorentz group action is naturally defined. Notice
that the $\cal P$ may be a submanifold of one of these spaces,
satisfying the requirements that all the Lorentz orbit belong to
this submanifold.

Two points are worth noting as we now transition from discussing the
theoretical frameworks to a first sketch of their phenomenological
implications. The two cases of positive and negative constant
curvature on $\cal P$ are expected to be rather different in their
phenomenological implications. Also,  while action of the lorentz
transformations will  be induced on the spacetime coordinates, this
action may be deformed and will depend on which coordinates are used
to label spacetime events and processes.  As we have seen the
induced spacetime coordinates are observer dependent, and are also
either non-commutative or energy-momentum dependent.  The details of
how these affect the action of lorent transformations on spacetime
reserved for discussion in a future publication.

\section{Determining experimentally the geometry of momentum space}

In the nineteenth century, Gauss proposed that the geometry of space
should  be empirically determined, and there is a legend he set
about measuring the curvature of space by determining the angles
between three mountain peaks.  Similarly, we here propose that the geometry of momentum space is to be measured rather than assumed.
In this section we  establish that the
geometry of momentum space can produce observable effects. We do
this by means of simple sketches of idealized  experiments.  We postpone for further work the
consideration of observations that could actually be done with present and near future technology.

There are two classes of experiments that can be contemplated.  There are experiments which aim to measure features of the geometry of
momentum space; these can be distinguished from tests of the specific hypotheses described in the previous section.
We describe two examples from the first class, which aim to measure the curvature of momentum space.

The first example we have chosen is from atomic physics, and shows that
the curvature of a connection on momentum space, defined by
(\ref{curvature1}) corresponds to a measurable quantity. We assume the
availability of a beam of atoms of same type prepared all in the
ground state with the same initial momentum. The idealized
measurement procedure we consider assumes the availability of four
energy levels: in addition to the ground state it involves excited
levels $I$, $II$ and $III$. Ideally we would want these energy
levels to be such that there  is a small step in energy from ground
to I excited, a large step in energy from $I$ to $II$ and another
small step in energy from $II$ to $III$. We excite the atoms with
lasers tuned to the transition between these states as shown in
Figure 2.

\begin{figure}[h!]
\begin{center}
\includegraphics[width=0.5 \textwidth]{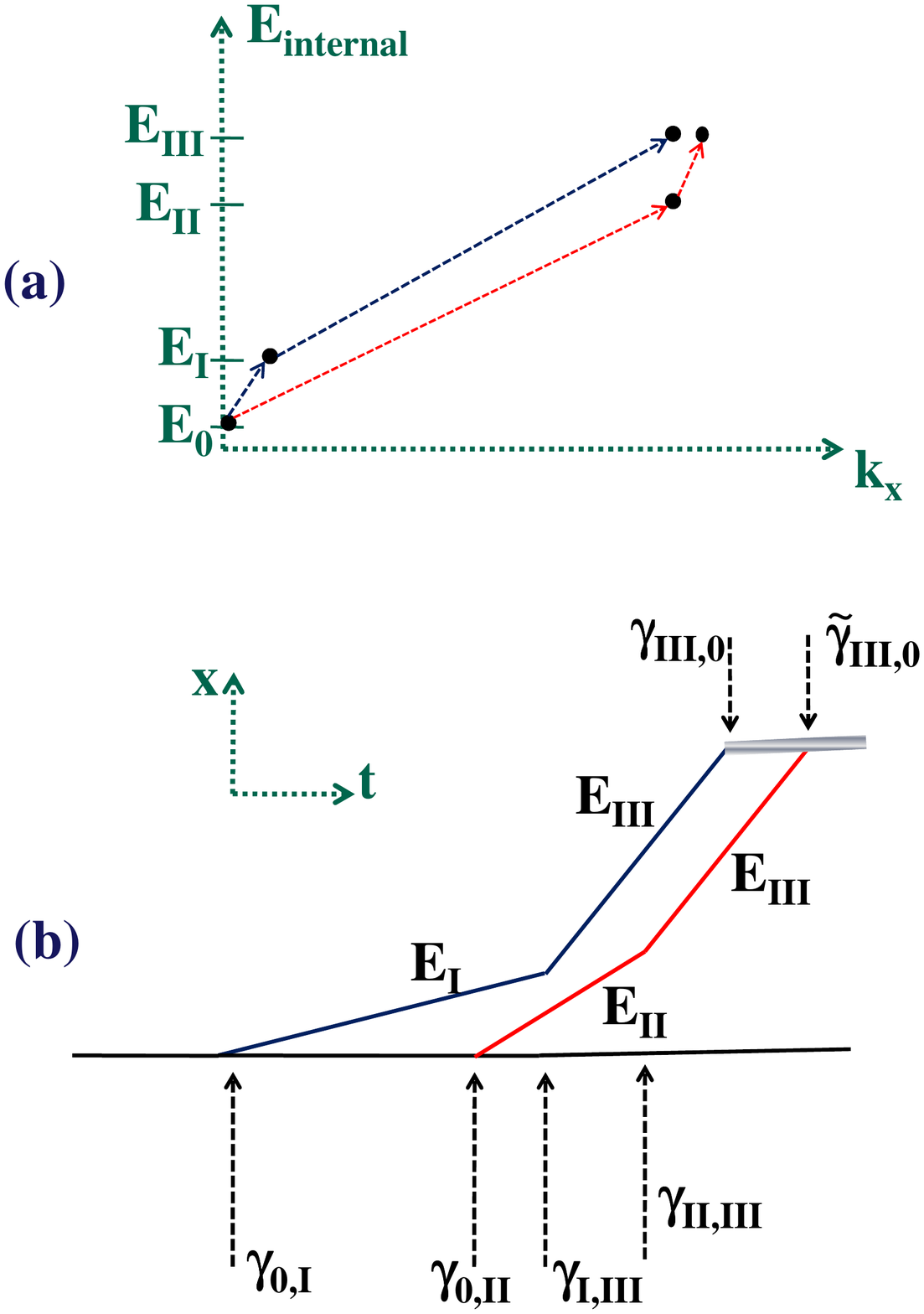}
\end{center}
\caption{Schematic description of a measurement procedure aimed at
exposing the implications of a non-trivial connection on momentum
space. The {\it routeA} and {\it routeB} described in the text are
here shown both in momentum space (panel (a)) and as spacetime
 trajectories of the atoms (panel (b)).}
\end{figure}

By the scheme in Figure 2 we can bring atoms from the ground state
to the level III excited  following two routes: \f route A  \ \
\equiv ground \rightarrow I \rightarrow III \ff and \f route B \ \
\equiv ground \rightarrow II  \rightarrow III \ff The resulting
momenta for the two processes are $(p \oplus k_{0,I})\oplus
k_{I,III}$ and $(p \oplus k_{0,II})\oplus k_{II,III}$, and can be
compared by measuring the laser frequency needed to bring the {\it
routeA} atoms back to the ground state and the laser frequency
needed to bring the {\it routeB} atoms back to the ground state. In
the idealized situation of  $k_{0,I}\simeq k_{II,III} \equiv q$,
$k_{0,II}\simeq k_{I,III} \equiv Q$, this procedure would give a
clean comparison between $(p \oplus q) \oplus Q$ and $(p \oplus
Q)\oplus q$, which following the analysis we offered in the previous
sections is indeed a sensitive indicator of the non-triviality of
the connection on momentum space. We expect that other more
practical and sensitive measurement procedures for the geometry of
momentum space will be gradually found, if a dedicated research
effort is inspired by our proposal. The simplicity of the scheme
described in Figure 2 serves the purpose of showing very clearly
that the geometry of momentum space can manifest itself in
measurable quantities.
 The procedure sketched in Figure 2
is also  representative of a whole class of strategies for measuring
non-associativity and/or non-commutativity of the law of composition of momenta,
which, as we have shown, are expressed respectively by the curvature and torsion of a connection on
$\cal P$.

It is interesting to note also that there is a simple analogy
between the non-linear composition of momenta we have discussed here
and the nonlinear law of composition of velocities, in Special
Relativity.  It is not always stressed that the composition law of
velocities in special relativity is  non-associative \cite{Girelli:2004xy}.  This
non-associativity is absent for addition of co-linear velocities but
is measured in Thomas procession \cite{Thomas:1926dy}.
This suggest an experiment inspired by the Thomas precession experiment \cite{Florian-etera-precession}.
The idea is to follow a system in orbit (an electron in an atom or  a particle circling in the LHC).
Such a system is enclosing a loop in momentum space at each period of revolution,
which enclose the curvature in momentum space.
At each period the localization of the orbiting particle will be shifted compared to the localization of a particle at rest.
Effectively, the particle will experience an infinitesimal boost $N_{i}$ at each period given by
\f
N_{i} =\frac{\Delta A_{cd}}{m_{P}^{2}} R^{cda}{}_{i} p_{a} \approx \frac{\Delta A_{cd}}{m_{P}^{2}} m R^{cd0}{}_{i}
\ff
where $\Delta A_{cd}$ is the area of the loop in momentum space and $m$ the mass of the particle.
One should be able to observe then a spacetime displacement due to relative locality effect.
Even if this effect is tiny this type of observation possess a huge potential since it is a cumulative effect and
 we can use the large number of orbits that develop over time.

 These experiments represent ways to measure the curvature of momentum space, which is the main effect that needs to be probed.
 We can also test  more broadly the solidity of the strong and weak dual equivalence principle.
 The metricity of the connection could be tested by looking for violations of the equivalence between mass and rest energy.
 There are numerous nuclear physics experiments that rely on the equivalence between mass and energy,
 the accuracy that is achieved in these experiments is not Planckian, but they clearly deserve a closer scrutiny in light of
 the new principle we propose.

 It will be also important to have a direct experimental bound on the momentum space torsion.
 One way to probe it is to put Bose statistics under experimental scrutiny,  since we have argued that a non vanishing torsion will in effect correspond to a modification of the statistics\footnote{The usual argument in favor of the standard statistics uses crucially the existence of an absolute spacetime, and the independence of the state of a system on its momentum space history. Logically, this derivation should be revisited in light of the relative locality principle}.
 This issue deserves a deeper analysis in order to propose specific effects to look for.
 Also one could imagine a momentum space E\"otvos experiment showing that all type of matter add momentum in the same way.

%A third kind of experiment would be precision measurements of processes where a particle at rest decays into a particle-antiparticle pair. These can be tests of  CPT symmetry, which is a well-tested symmetry~\cite{kloe2} establish that two particles on shell (and onshellness is established by the  momentum-space metric within our framework) can have total momentum with vanishing spatial component, $(p_1\oplus p_2)_j = 0$, and the zero value for the spatial component of spatial momentum must taken into account, within our framework, of the connection in momentum space.

We can add that a non-trivial geometry of momentum space can in the most general form
produce deformations of well tested symmetries of quantum field
theory including $CPT$ and crossing symmetry\cite{CPT-Giovanni}.  This is because they imply non-linearities in
conservation laws, which would show up as violations of the linear form of those laws.   For example, we expect that standard
arguments on crossing symmetry in which an incoming particle with four momentum $k_a$ is replaced
by an outgoing antiparticle with four momentum $-k_a$ will be deformed so the antiparticle has
instead momentum $\ominus k_a$.  This will introduce non-linearities which will show up as violations
of the usual crossing symmetry.   Thus the tight experimental
constraints on these symmetries constrains the geometry of momentum space.  This is expressed by the
{\it correspondence principle} we discussed in the previous section.
Thus, a first task for phenomenologists will be
to understand the bounds on the mass scale $m_c$ from existing tests of fundamental symmetries in quantum
field theory \cite{CPT-Giovanni}.

These examples illustrates that the principles  proposed here open a
new type of investigation, both experimental and theoretical, into
the geometry of momentum space.

\section{Conclusions}

The passage from special relativistic locality to relative locality
reminds us  of the passage we navigated a century ago from absolute
space to spacetime. The additivity of velocity implies there is an
absolute time by which velocity is measured.  If we hypothesize that
the combinations of velocity might become non-linear, without
weakening the principle of the relativity of inertial frames, we
need an invariant scale, to measure the scale of the
non-linearities, which must be a velocity itself.  Hence there is an
invariant  velocity we call $c$.  This then  allows us to
interchange distances and times, which makes possible the existence
of an absolute spacetime, which replaces the notion of absolute
space.  Space itself remains, but as an observer dependent concept,
because of the relativity of the simultaneity of distant events.

Similarly, as we observed above, the additivity of momenta and
energy  implies the existence of an absolute spacetime.  When we
contemplate weakening that to a non-linear combination rule for
momenta in physical interactions, we need an invariant momentum
scale.   We have taken this scale  to be $m_p$ but of course from a
phenomenological point of view it should be taken as having a free
value to be constrained by experiment.   This, together with $\hbar$
makes it possible to interchange distances and momenta, which makes
possible the mixing of spacetime coordinates with energy and
momenta, so that the only invariant structure is the phase space. We
saw above explicitly how non-linearity in conservation of energy and
momentum directly forces translations of spacetime coordinates to
depend on momenta and energies.  Local spacetime remains, but as an
observer dependent concept, because of the relativity of the
locality of distant events.

Relative locality suggests novel points of departure for attempts to
discover the right quantum theory of gravity.  Here, we have
discussed  its implications in a semi-phenomenological perspective
framed by the approximation (\ref{regime}) but it is possible it
goes deeper. For example, one can suggest that the fundamental
description is one in which there is dynamical curvature in phase
space, so that the fundamental constants are $G_{Newton}$ and $m_p$.
This implies that Planck's constant could be a derived quantity,
$\hbar=G_{Newton} m^2_p$, suggesting that quantum mechanics is
emergent from a dynamics of phase space geometry.

The idea that momentum space might be curved has a long history.
To our knowledge, it was first put forward more than 70 years ago in
a paper by Max Born \cite{Born1938}.  The idea seems to have been
next discussed independently
 in the celebrated paper by Snyder \cite{Snyder:1946qz}, which was
 followed by series of papers of Russian physicists from the 1950s to the 1980s \cite{golfand}.
 There the main motivation was the ultraviolet divergencies in quantum field theories;
  it was hoped that curving momentum space may help taming these divergencies.

As Snyder first pointed out, the curvature of momentum space
implies the non-commutativity of spacetime coordinates which
generate translations on momentum space.  A picture of quantum
geometry in which a curved momentum space is dual to a
non-commutative spacetime was introduced by Majid, who stressed the
relation to quantum groups \cite{Majid:1999tc}.  This idea was
realized in the construction of $\kappa$-Minkowski spacetime which
is a non-commutative geometry invariant under a deformation of
Poincare symmetry known as the $\kappa$-Poincare algebra
\cite{Lukierski:1991pn}, \cite{Majid:1994cy}\footnote{The geometry
of $\kappa$-Poincare theory fits the general scheme presented here
with a connection with non-zero torsion but zero curvature.}.

Still more generally, Alain Connes has
stressed~\cite{connesMOMENTUM} that  the fundamental observables are
spectra of energy and momentum, tightly cementing the relationship
between the geometry of momentum space and non-commutative geometry.

Another motivation comes from lower dimensional physics. It is by
now well  established by different methods
\cite{Matschull:1997du,Bais:1998yn,Bais:2002ye,Meusburger:2003ta,AmelinoCamelia:2003xp,Freidel:2003sp,Freidel:2005me,Schroers:2007ey}
that in $2+1$ dimensions, where gravity is described by a
topological field theory \cite{Witten:1988hc}, the effective
momentum space of particles is deformed and becomes a (curved) three
dimensional group manifold.

So the world we have described here is realized in at least one well
developed example, which is quantum gravity coupled to matter in
$2+1$ dimensions.

In the physical $3+1$ dimensions the situation is less clear. It has
been suggested that, as in the 3D case, the ``no gravity'' limit is
governed by topological field theory, effectively leading to curved
momentum space \cite{KowalskiGlikman:2008fj}. The curved momentum
space was also employed in the context of group field theory, which
generates amplitudes for spin foam models of quantum gravity
\cite{Oriti:2009wn}. Quantum field theory with curved momentum space
has been recently discussed in detail in \cite{Arzano:2010jw}. The
dynamical momentum space in the context of the cosmological constant
problem has been recently discussed \cite{Chang:2010ir}.

Last but not least, momentum spaces of constant curvature find  its
natural application as a model of Doubly (or deformed) Special
Relativity (DSR)
\cite{Amelino-Camelia:2000ge,Amelino-Camelia:2000mn,Kowalski-Glikman:2002ft,Magueijo:2001cr},
whose second observer independent scale is naturally associated with
the curvature of the momentum space
\cite{Kowalski-Glikman:2002ft}\footnote{The existence of a regime
defined by eq.\ (\ref{regime}) was discussed by several authors
including \cite{majidREGIMES,KowalskiGlikman:2003we,carloREGIMES}.} (see also \cite{Moffat:2004jj}.)
These formulations were closely related to non-commutative geometry.
A second formulation of DSR expressed the same idea as an energy
dependence of spacetime \cite{Magueijo:2001cr}.  For a long time it
has been suspected that these were different ways of formulating the
same theories, the developments of this paper show how an energy
dependent metric and non-commutative spacetime coordinates are
different ways of expressing a deeper idea, which is relative
locality.

The non-associativity of momentum addition in non-commutative
geometry was also explored in \cite{Girelli:2004xy}, which stressed
the analogy to the non-associativity of velocity composition in
special relativity. The non-associativity of momentum conservation
was also further explored in  \cite{Girelli:2010zw}\footnote{While
finishing this paper we learned that our construction is related to
the mathematical theory of loops developed, among others, by M.\
Kikkawa and L.V.\ Sabinin, \cite{KS}.  We thank F.\ Girelli and E.\
Livine for pointing this out to us.}.

The issue of possible macroscopic non-local effects in DSR was
raised several times
\cite{gacIJMPdsrREV,dedeo,Schutzhold:2003yp,Arzano:2003da,Hossenfelder:2010tm}.
Attempts to address these issues led to some partial anticipations
of the ideas proposed here \cite{whataboutbob,leeINERTIALlimit,JM1}.
From the present perspective we can appreciate that the concern was
well justified and there is a pot of gold under the rainbow of
apparent non-locality.  At the same time, the principle of relative
locality tells us that the theories discussed here do not contain
physical non-localities of the kind that were suggested in
\cite{dedeo,Schutzhold:2003yp,Hossenfelder:2010tm}.  Instead,
energy-momentum conserving interactions are always local in
spacetime when observed by observers who are local to them. The fact
that a distant observer is only able to localize an event involving
several particles with different momenta to within a region whose
scale is proportional to its distance is inevitable in theories with
curved momentum spaces, and should not be misconstrued as entailing
a violation of the  physical principle that  physical interactions
are local.

Nonetheless, if theories are still local, in the restricted sense of
relative locality, there are, as we discussed above, new phenomena
that can be studied experimentally.  In addition, as will be
explained elsewhere, the problem of whether photons emitted
simultaneously are detected simultaneously, after traveling for very
long distances, can be cleanly addressed.  This is relevant for the
timing of arrival time measurements in gamma ray
bursts~\cite{ellis09,unoEdue,fermiNATURE}.

 We have seen here that the notion of curved momentum space  has generic and
 vivid consequences for our understanding of basic physics.  We do not live in spacetime.
 We live in Hilbert space, and the classical approximation to that is that we live in phase space.
 If we think that we can filter out colors and frequencies of particles to arrive at a picture of
 particles moving in spacetime which is independent of the momenta those particles carry,
 that is only an illusion that has been possible because of the smallness of elementary
 particle scales in Planck units.  Similarly, if we think that observers distant
 from each other see the same spacetime that is only because we are only beginning to make precise measurements of quantities where the combination
 $|z||\Gamma | k|/\hbar $ is not negligible.

 The crucial idea underlying and unifying all these developments turns
 out to be one that is a direct consequence of the curvature of momentum space: the relativity of locality.

Even apart from fundamental physics, there are situations in
condensed matter physics, where it is convenient to understand
excitations as living in a curved momentum space\cite{ryu}.
The considerations of this paper may be relevant for those cases.
Or, to put it the other way, just as some condensed matter or fluid
systems provide analogues for relativity and gravity, it may be that
condensed matter systems with curved momentum spaces may give us analogues to the physics of relative locality.

So look around.  You see colors and angles, {\it i.e.}  you are
seeing into phase space.  The idea that underlying it is an energy
independent,  invariant spacetime geometry could be an
approximation, reliable only to the extent that we measure the
geometry with quanta small compared to the Planck energy and we
neglect phenomena of order of $|z||\Gamma | k|/\hbar $.  Whether
this is correct or not is for experimental physics to decide. If it
turns out to be correct, then a new arena opens up for experimental
physics and astronomy, which is the measurement of the geometry of
momentum space.

\section*{ACKNOWLEDGEMENTS}

We are very grateful to Stephon Alexander, Michele Arzano, James
Bjorken, Florian Girelli, Sabine Hossenfelder, Viqar Husain, Etera
Livine, Seth Major, Djorje Minic, Carlo Rovelli, Frederic Schuller,
Guglielmo Tino, and William Unruh for conversations and
encouragement.  GAC and JKG thank Perimeter Institute for
hospitality during their visits in September 2010, when the main
idea of this paper was conceived.  The work of G. Amelino-Camelia
was supported in part by grant RFP2-08-02 from The Foundational
Questions Institute (fqxi.org). The work of J. Kowalski-Glikman
was supported in part by grants 182/N-QGG/2008/0 and N 202331139.
  Research at Perimeter Institute
for Theoretical Physics is supported in part by the Government of
Canada through NSERC and by the Province of Ontario through MRI.


\begin{thebibliography}{99}

\bibitem{AE-SR}A. Einstein, {\it Zur Elektrodynamik bewegter K\"orper}, Annalen der Physik {\bf 17}, 891 (1905)  English translation {\it On the electrodynamics of moving bodies},  in {\it The principle of relativity: a collection of original memoirs}
 By Hendrik Antoon Lorentz, Albert Einstein, H. Minkowski, Hermann Weyl, Dover books.
 (English translation also at
http://www.fourmilab.ch/etexts/einstein/specrel/specrel.pdf).

\bibitem{Laurent-math}L. Freidel, {\it The geometry of momentum space}, preprint in preparation.

\bibitem{E=MC2}L. Smolin et al, preprint in preparation.

\bibitem{CPT-Giovanni}G. Amelino-Camelia et. al, preprint in preparation.

\bibitem{Girelli:2004xy}
  F.~Girelli and E.~R.~Livine,
 {\it  Special Relativity as a non commutative geometry: Lessons for Deformed
  Special Relativity,}
  Phys.\ Rev.\  D {\bf 81} (2010) 085041
  [arXiv:gr-qc/0407098].

\bibitem{Thomas:1926dy}
  L.~H.~Thomas,
  {\it  The motion of a spinning electron,}
  Nature {\bf 117}, 514 (1926);
  ibid.\ {\it  The Kinematics Of An Electron With An Axis,}
  Phil.\ Mag.\  {\bf 3}, 1 (1927). E.~P.~Wigner,
  {\it  On Unitary Representations Of The Inhomogeneous Lorentz Group,}
  Annals Math.\  {\bf 40}, 149 (1939)
  [Nucl.\ Phys.\ Proc.\ Suppl.\  {\bf 6}, 9 (1989)].

  \bibitem{Florian-etera-precession}F. Girelli, E.R. Livine, {\it Physics of Deformed Special Relativity: Principle of Relativity revisited}, gr-
qc/0412004; {\it  Deformed Special Relativity: problems and solutions} , Braz. J. Phys. 35 (2005) 432-
438, gr-qc/0412079.


\bibitem{Born1938}Max Born, {\it A Suggestion for Unifying Quantum Theory and Relativity}, Proc. R. Soc. Lond. A, 1938 {\bf 165}
291-303.

\bibitem{Snyder:1946qz}
  H.~S.~Snyder,{\it  Quantized space-time,}
  Phys.\ Rev.\  {\bf 71}, 38 (1947).

\bibitem{golfand}
Yu. A. Gol'fand,
{\it  On the introduction of an ``elementary length" in the relativistic theory of elementary particle},
JETP {\bf 37} (1959) 504-509; English in: Soviet Physics JETP {\bf 10} (1960) 356-360; V. G. Kadyshevskii,
{\it  On the theory of quantization of space-time},
JETP {\bf 41} (1961) 1885-1894, English in: Soviet Physics JETP {\bf 14} (1962) 1340-1346; Yu. A. Gol'fand,
{\it  Quantum field theory in constant curvature p-spac},
JETP {\bf 43} (1962) 256-257, English in: Soviet Physics JETP {\bf 16} (1963) 184-191; V.~G.~Kadyshevsky, M.~D.~Mateev, R.~M.~Mir-Kasimov and I.~P.~Volobuev,
 {\it  Equations Of Motion For The Scalar And The Spinor Fields In
  Four-Dimensional Noneuclidean Momentum Space,}
  Theor.\ Math.\ Phys.\  {\bf 40} (1979) 800
  [Teor.\ Mat.\ Fiz.\  {\bf 40} (1979) 363]; V.~G.~Kadyshevsky and M.~D.~Mateev,
  {\it  Quantum Field Theory And A New Universal High-Energy Scale: The Scalar
  Model,}
  Nuovo Cim.\  A {\bf 87} (1985) 324.

\bibitem{Majid:1999tc}
  S.~Majid,
  {\it  Meaning of noncommutative geometry and the Planck-scale quantum group,}
  Lect.\ Notes Phys.\  {\bf 541} (2000) 227
  [arXiv:hep-th/0006166].

  %\cite{Lukierski:1991pn}
\bibitem{Lukierski:1991pn}
  J.~Lukierski, H.~Ruegg, A.~Nowicki {\it et al.},
  {\it Q deformation of Poincare algebra,}
  Phys.\ Lett.\  {\bf B264 } (1991)  331-338.

  %\cite{Majid:1994cy}
\bibitem{Majid:1994cy}
  S.~Majid, H.~Ruegg,
   {\it Bicrossproduct structure of kappa Poincare group and noncommutative geometry,}
  Phys.\ Lett.\  {\bf B334 } (1994)  348-354.
  [hep-th/9405107].

\bibitem{connesMOMENTUM}A. Connes,  {\it Non-commutative geometry} Academic Press, 1994.

%\cite{Matschull:1997du}
\bibitem{Matschull:1997du}
  H.~-J.~Matschull, M.~Welling,
   {\it Quantum mechanics of a point particle in (2+1)-dimensional gravity,}
  Class.\ Quant.\ Grav.\  {\bf 15 } (1998)  2981-3030.
  [gr-qc/9708054].

  %\cite{Bais:1998yn}
\bibitem{Bais:1998yn}
  F.~A.~Bais, N.~M.~Muller,
   {\it Topological field theory and the quantum double of SU(2),}
  Nucl.\ Phys.\  {\bf B530 } (1998)  349-400.
  [hep-th/9804130].

  %\cite{Bais:2002ye}
\bibitem{Bais:2002ye}
  F.~A.~Bais, N.~M.~Muller, B.~J.~Schroers,
   {\it Quantum group symmetry and particle scattering in (2+1)-dimensional quantum gravity,}
  Nucl.\ Phys.\  {\bf B640 } (2002)  3-45.
  [hep-th/0205021].

  %\cite{Meusburger:2003ta}
\bibitem{Meusburger:2003ta}
  C.~Meusburger and B.~J.~Schroers,
   {\it Poisson structure and symmetry in the Chern-Simons formulation of
  (2+1)-dimensional gravity,}
  Class.\ Quant.\ Grav.\  {\bf 20} (2003) 2193
  [arXiv:gr-qc/0301108].
  %%CITATION = CQGRD,20,2193;%%


  %\cite{AmelinoCamelia:2003xp}
\bibitem{AmelinoCamelia:2003xp}
  G.~Amelino-Camelia, L.~Smolin and A.~Starodubtsev,
   {\it Quantum symmetry, the cosmological constant and Planck scale
  phenomenology,}
  Class.\ Quant.\ Grav.\  {\bf 21} (2004) 3095
  [arXiv:hep-th/0306134].
  %%CITATION = CQGRD,21,3095;%%

%\cite{Freidel:2003sp}
\bibitem{Freidel:2003sp}
  L.~Freidel, J.~Kowalski-Glikman and L.~Smolin,
   {\it 2+1 gravity and doubly special relativity,}
  Phys.\ Rev.\  D {\bf 69} (2004) 044001
  [arXiv:hep-th/0307085].
  %%CITATION = PHRVA,D69,044001;%%

  %\cite{Freidel:2005me}
\bibitem{Freidel:2005me}
 L.~Freidel, E.~R.~Livine,
   {\it Ponzano-Regge model revisited III: Feynman diagrams and effective field theory,}
  Class.\ Quant.\ Grav.\  {\bf 23}, 2021-2062 (2006).
  [hep-th/0502106].\\
  L.~Freidel and E.~R.~Livine,
   {\it Effective 3d quantum gravity and non-commutative quantum field theory,}
  Phys.\ Rev.\ Lett.\  {\bf 96} (2006) 221301
  [arXiv:hep-th/0512113].
  %%CITATION = PRLTA,96,221301;%%

%\cite{Schroers:2007ey}
\bibitem{Schroers:2007ey}
  B.~J.~Schroers,
   {\it Lessons from (2+1)-dimensional quantum gravity,}
  PoS {\bf QG-PH} (2007) 035
  [arXiv:0710.5844 [gr-qc]].
  %%CITATION = POSCI,QG-PH,035;%%

%\cite{Witten:1988hc}
\bibitem{Witten:1988hc}
  E.~Witten,
   {\it (2+1)-Dimensional Gravity as an Exactly Soluble System,}
  Nucl.\ Phys.\  B {\bf 311} (1988) 46.
  %%CITATION = NUPHA,B311,46;%%

%\cite{KowalskiGlikman:2008fj}
\bibitem{KowalskiGlikman:2008fj}
  J.~Kowalski-Glikman and A.~Starodubtsev,
   {\it Effective particle kinematics from Quantum Gravity,}
  Phys.\ Rev.\  D {\bf 78} (2008) 084039
  [arXiv:0808.2613 [gr-qc]].
  %%CITATION = PHRVA,D78,084039;%%

%\bibitem{Arzano:2010kz} M.~Arzano and J.~Kowalski-Glikman,  {\it  Kinematics of a relativistic particle with de Sitter momentum space,}   arXiv:1008.2962 [hep-th].

%\cite{Oriti:2009wn}
\bibitem{Oriti:2009wn}
 L.~Freidel,
   {\it Group field theory: An Overview,}
  Int.\ J.\ Theor.\ Phys.\  {\bf 44}, 1769-1783 (2005).
  [hep-th/0505016].\\
  D.~Oriti,  {\it  The group field theory approach to quantum gravity: some recent results,}
  arXiv:0912.2441 [hep-th].


\bibitem{Arzano:2010jw}
  M.~Arzano,  {\it Anatomy of a deformed symmetry: field quantization on curved momentum
  space,}
  arXiv:1009.1097 [hep-th].

  \bibitem{Chang:2010ir}
  L.~N.~Chang, D.~Minic, T.~Takeuchi,
  {\it Quantum Gravity, Dynamical Energy-Momentum Space and Vacuum
  Energy,}
  Mod.\ Phys.\ Lett.\  {\bf A25 } (2010)  2947-2954.
  [arXiv:1004.4220 [hep-th]].


\bibitem{Amelino-Camelia:2000ge}
G.~Amelino-Camelia,  {\it Testable scenario for relativity with
minimum-length,}  Phys.\ Lett.\ B {\bf 510}, 255 (2001)
[arXiv:hep-th/0012238].

\bibitem{Amelino-Camelia:2000mn}
G.~Amelino-Camelia,  {\it Relativity in space-times with short-distance
structure governed by an observer-independent (Planckian) length
scale,}  Int.\ J.\ Mod.\ Phys.\ D {\bf 11}, 35 (2002)
[arXiv:gr-qc/0012051].


\bibitem{Kowalski-Glikman:2002ft}
J.~Kowalski-Glikman,  {\it  De Sitter space as an arena for doubly
special relativity,} Phys.\ Lett.\ B {\bf 547} (2002) 291
[arXiv:hep-th/0207279].

 %\cite{Moffat:2004jj}
\bibitem{Moffat:2004jj}
  J.~W.~Moffat,
  ``Quantum gravity momentum representation and maximum invariant energy,''
  arXiv:gr-qc/0401117.
  %%CITATION = GR-QC/0401117;%%

\bibitem{Magueijo:2001cr}
  J.~Magueijo, L.~Smolin,
   {\it Lorentz invariance with an invariant energy scale}
  Phys.\ Rev.\ Lett.\  {\bf 88 } (2002)  190403.
  [hep-th/0112090]; {\it   Generalized Lorentz invariance with an invariant energy scale },  gr-qc/0207085,  Phys.Rev. D67 (2003) 044017.

  \bibitem{Girelli:2010zw}
  F.~Girelli,
   {\it Snyder Space-Time: K-Loop and Lie Triple System,}
  SIGMA {\bf 6}, 074 (2010).
  [arXiv:1009.4762 [math-ph]].

\bibitem{gacIJMPdsrREV}
G.~Amelino-Camelia,
  {\it  Doubly special relativity: First results and key open problems}
Int.\ J.\ Mod.\ Phys.\ \textbf{D11} (2002) 1643
[arXiv:gr-qc/0210063].  (See the paragraph before eq. 21).

\bibitem{dedeo}
S.~DeDeo and C.~Prescod-Weinstein,
  {\it Energy-Dependent Speeds of Light for Cosmic-Ray Observatories}
arXiv:0811.1999.

  %\cite{Schutzhold:2003yp}
\bibitem{Schutzhold:2003yp}
  R.~Schutzhold, W.~G.~Unruh,
   {\it Large-scale nonlocality in 'doubly special relativity' with an energy-dependent speed of light,}
  JETP Lett.\  {\bf 78 } (2003)  431-435.
  [gr-qc/0308049].


  %\cite{Hossenfelder:2010tm}
\bibitem{Hossenfelder:2010tm}
  S.~Hossenfelder,
  {\it Bounds on an energy-dependent and observer-independent speed of light from violations of locality,}
  Phys.\ Rev.\ Lett.\  {\bf 104 } (2010)  140402.
  [arXiv:1004.0418 [hep-ph]].


  %\cite{Arzano:2003da}
\bibitem{Arzano:2003da}
  M.~Arzano,
  {\it Comment on 'Large scale nonlocality in 'doubly special relativity' with an energy dependent speed of light',}
  [gr-qc/0309077].

\bibitem{whataboutbob}
  G.~Amelino-Camelia, M.~Matassa, F.~Mercati and G.~Rosati,
 {\it Taming nonlocality in theories with deformed Poincare symmetry,}
  arXiv:1006.2126 [gr-qc];  U.~Jacob, F.~Mercati, G.~Amelino-Camelia and T.~Piran,
  {\it Modifications to Lorentz invariant dispersion in relatively
    boosted frames,}
  Phys.\ Rev.\  D {\bf 82} (2010) 084021
  [arXiv:1004.0575 [astro-ph.HE]]

\bibitem{leeINERTIALlimit}
L.~Smolin,{\it On limitations of the extent of inertial frames in
non-commutative relativistic spacetimes},  arXiv:1007.0718;    {\it
Classical paradoxes of locality and their possible quantum
resolutions in deformed special relativity}, arXiv:1004.0664

\bibitem{JM1}M. Arzano, J. Kowalski-Glikman {\it Kinematics of a relativistic particle with de Sitter momentum space}, arXiv:1008.2962 [hep-th].

%\cite{KowalskiGlikman:2003we}
\bibitem{KowalskiGlikman:2003we}
  J.~Kowalski-Glikman, S.~Nowak,
   {\it Doubly special relativity and de Sitter space,}
  Class.\ Quant.\ Grav.\  {\bf 20 } (2003)  4799-4816.
  [hep-th/0304101].



\bibitem{ellis09} John Ellis, N.E. Mavromatos, D.V. Nanopoulos,
  {\it Probing a Possible Vacuum Refractive Index with Gamma-Ray Telescopes}
 Phys.~Lett.~B674 (2009) 83. [arXiv:0901.4052]

\bibitem{unoEdue}
  G.~Amelino-Camelia and L.~Smolin,
  {\it Prospects for constraining quantum gravity dispersion with near term
  observations,}
  Phys.\ Rev.\  D {\bf 80} (2009) 084017
  [arXiv:0906.3731 [astro-ph.HE]].
  %%CITATION = PHRVA,D80,084017;%%

\bibitem{fermiNATURE}
A. A. Abdo et al,
%. [Fermi LAT/GBM Collaborations],
  {\it A limit on the variation of the speed of light arising from quantum gravity effects}
%Nature \textbf{462}, 331--334 (2009).
Nature \textbf{462} (2009) 331.

\bibitem{KS}Kikkawa M.,  {\it  Geometry of homogeneous Lie loops}, Hiroshima Math. J. 5
(1975), 141-179;  Sabinin L.V.,  {\it Smooth quasigroups and loops},
Kluwer Academic Publishers, Dordrecht, 1999

  \bibitem{majidREGIMES}
  S.~Majid and R.~Oeckl, {\it Twisting of quantum differentials and the Planck scale Hopf algebra,}
  Commun.\ Math.\ Phys.\  {\bf 205} (1999) 617
  [arXiv:math/9811054].

\bibitem{carloREGIMES}
  C.~Rovelli,
  {\it A note on DSR,}
  arXiv:0808.3505 [gr-qc].





\bibitem{ryu}Shunji Matsuura, Shinsei Ryu, {\it Momentum space metric, non-local operator, and topological insulators},
 arXiv:1007.2200.

\end{thebibliography}
\end{document}